\documentclass[12pt]{article}
\newif\ifpdf\ifx\pdfoutput\undefined\pdffalse\else\pdfoutput=1\pdftrue\fi
\newcommand{\pdfgraphics}{\ifpdf\DeclareGraphicsExtensions{.pdf,.jpg}
\else\fi} 
\usepackage{graphicx,epstopdf,amssymb,amsfonts,amsmath,amsthm,array,
mathrsfs,amscd}
\newcommand{\pbs}[1]{\let\temp=\\#1\let\\=\temp}
\numberwithin{equation}{section}
\def\be{\begin{equation}}\def\ee{\end{equation}}
\def\cvp{\raise 2pt\hbox{,}} 
 
  \def\Tr{\mathop{\rm Tr}\nolimits}

 \def\nn{{\cal
N}}

 \def\uN{{\text U}(N)}  \def\M{\mathscr M} \def\u{\text{U}(1)} \def\SN{\text{S}_{N}}

\def\La{\Lambda}
\def\g{\boldsymbol{g}}\def\q{\boldsymbol{q}} \def\ring{\mathsf A}\def\field{\mathsf
K}\def\pfield{\mathsf k} \def\pring{\mathsf a}\def\ideal{\mathscr I}
\def\fieldp{\field_{|\varphi)}}
\def\vp{|\varphi)}\def\split{\mathsf S}

\theoremstyle{plain}
\newtheorem{thm}{Theorem}
\newtheorem{lem}[thm]{Lemma}

\theoremstyle{definition}
\theoremstyle{remark}

\def\plb#1#2#3{{\it Phys.\ Lett.\ }{\bf B #1} (#2) #3}
\def\npb#1#2#3{{\it Nucl.\ Phys.\ }{\bf B #1} (#2) #3}

\def\jhep#1#2#3{{\it J. High Energy Phys.\ }{\bf #1} (#2) #3}
\def\prd#1#2#3{{\it Phys.\ Rev.\ }{\bf D #1} (#2) #3}

\begin{document}
\pdfgraphics 
%
\pagestyle{empty}
{\parskip 0in

\hfill LPTENS-08/54

\vfill
\begin{center}
{\LARGE Galois Symmetries in Super Yang-Mills Theories}



\vspace{0.4in}

Frank F{\scshape errari}
\\
\medskip
{\it Service de Physique Th\'eorique et Math\'ematique\\
Universit\'e Libre de Bruxelles and International Solvay Institutes\\
Campus de la Plaine, CP 231, B-1050 Bruxelles, Belgique\\
\smallskip
and\\ \smallskip
Simons Center for Geometry and Physics\\
Stony Brook University, Stony Brook, NY 11794-3840, USA\\}
\smallskip
{\tt frank.ferrari@ulb.ac.be}
\end{center}
\vfill\noindent

Classifying the phases of gauge theories is hindered by the lack of
local order parameters. In particular, the standard Wilson's and
't~Hooft's non-local order parameters are known to be insufficient to
explain the existence of the plethora of phases that are found in
supersymmetric gauge theories. Motivated by these observations, we
reanalyze the concept of gauge symmetry breaking using Galois theory.
Unlike the ordinary classical notion of unbroken gauge group, the
Galois symmetry makes sense in the full quantum theory and must be a
phase invariant. The algebraic structure underlying the space of vacua
of supersymmetric gauge theories, that we have developed recently, is
precisely designed to allow a rigorous mathematical implementation of
these ideas.

\vfill

\medskip
%
\begin{flushleft}
\today
\end{flushleft}
\newpage\pagestyle{plain}
\baselineskip 16pt
\setcounter{footnote}{0}

%
\section{Introduction}
\label{IntroSec}

The fundamental idea, due to Landau, to understand and classify the
possible phases of a physical system is based on the concept of
symmetry. Suppose that the physics is invariant under some symmetry
group in some regime. Since the group structure is rigid, it cannot
change when the parameters of the theory are varied, except by going
through a singularity, which corresponds to a phase transition. The
symmetry group can thus be used to characterize the phases. In
particular, if in some regime the symmetry group is $E$ and in another
regime it is $E'\not = E$, then a smooth interpolation between the two
regimes cannot exist: they must be in different phases. For example,
it is not possible to go smoothly from solid to liquid water (the
liquid has full translational symmetry whereas in the solid this
symmetry is broken down to a discrete subgroup), whereas the
transition from liquid to vapour can be smooth (this is associated
with the existence of the critical point in the water phase diagram).

The symmetry $E$ can be determined by computing the expectation values
of observables, called order parameters in this context, that
transform non-trivially under $E$. One of the main interest in
Landau's ideas is that this computation can be done \emph{locally} in
parameter space, i.e.\ by looking at the system in some particular
regime, without needing to probe the full phase diagram. For example,
the translational symmetry of the liquid/vapour phase of water can be
seen by looking at a regime of pressure and temperature where water
looks like a liquid, or like vapour, but one doesn't need to look at
both. The fact that they belong to the same phase ensures that the
result for $E$ will be the same in both regimes. In other words, the
symmetry $E$ is a \emph{phase invariant}. It is a very powerful tool,
because from a local analysis one obtains global constraints on the
phase diagram.

One would like to apply the same ideas to classify the phases of gauge
theories. However, in this case, one runs into well-known subtleties.
The basic difficulty is that gauge invariance is not really a
symmetry. It is more accurately described as a redundancy in the
description of the system under consideration. Physical observables
must be gauge invariant and thus we do not see the gauge group, at
least in any obvious way, in the physics. In particular, there is no
obstacle in principle in having equivalent physical theories based on
two different gauge groups, or in having vacua belonging to the same
phase but having different patterns of gauge symmetry breaking. There
is also no obstacle in principle in having a formulation of the theory
based only on physical observables, in which the original notion of
gauge invariance is altogether absent. Actually, this is exactly what
happens in the closed string description of gauge theories.

To be a little bit more precise, let us consider a four dimensional
gauge theory based on a compact Lie group $G$ and let us denote by
$\mathscr G$ the group of local gauge transformations. As a
consequence of Gauss' law, the theory is defined by modding out by the
group $\mathscr G_{0}$ of gauge transformations that are connected to
the identity. One is thus left in principle with a symmetry group
$\mathscr G/\mathscr G_{0}=G_{\infty}\simeq G$ which is often called
``the group of gauge transformations at infinity.'' Imagine now that
you consider a local observable $\mathcal O(x)$. As for any observable
in the theory, it must be invariant under $\mathscr G_{0}$. But for a
\emph{local} operator, invariance under $\mathscr G_{0}$ obviously
implies invariance under $\mathscr G$ and thus $G_{\infty}$ as well.
Thus we see that in gauge theory, \emph{there is no local order
parameter}, at least in any obvious sense.\footnote{Of course gauge
theories may have global symmetries and associated local order
parameters. We are focusing in this paper on the possibility to use
the gauge symmetry itself in order to distinguish the phases and thus 
in particular to understand the phase structure of gauge systems that 
do not have a global symmetry.}

On the other hand, one can use non-local order parameters, like the
Polyakov-'t~Hooft-Wilson's lines. This yields many interesting
results. For example, it is widely believed that the so-called
't~Hooft's classification of the \emph{massive} phases of gauge
theories that one obtains in this way is complete (see for example
\cite{DWitten} for a nice discussion and references). Unfortunately,
for phases with no mass gap, 't~Hooft's arguments are not powerful
enough to provide a complete classification. These massless phases are
ubiquitous in gauge theories, since they occur for instance each time
there is a massless photon in the spectrum.

Our goal in the present paper is to try to shed some new light on this
problem by using the nice mathematical structures underlying the
solutions of $\nn=1$ supersymmetric gauge theories \cite{ferirr}. We
shall see that the concept of gauge symmetry breaking which, in its
usual form, makes sense only classically, can be elevated to a
well-defined quantum concept using Galois theory. The Galois symmetry
is shown to be a phase invariant and thus can be used to classify the
phases, in line with Landau's ideas. The Galois group is also an
intrinsic property of the phase, independent of the particular
realization of the phase in a given model.

For concreteness, we shall focus on the paradigmatic example of the
$\nn=1$ theory with gauge group $\uN$, one adjoint chiral multiplet
$\phi$ and tree-level superpotential $\Tr W(\phi)$ such that
\be\label{Wtree} W'(z) = \sum_{k=0}^{d}g_{k} z^{k} =
g_{d}\prod_{i=1}^{d}(z-w_{i})\, .\ee
As is well known, the vacua of this model can be labeled as
$|N_{1},k_{1};\ldots;N_{d},k_{d}\rangle$, where $0\leq k_{i}\leq
N_{i}-1$. The integer $N_{i}$ corresponds to the number of eigenvalues
of $\phi$ that are equal to $w_{i}$ classically. The pattern of gauge
symmetry breaking in $|N_{1},k_{1};\ldots;N_{d},k_{d}\rangle$ is thus
$\uN\rightarrow\text{U}(N_{1})\times\cdots\times\text{U}(N_{d})$. The
integer $k_{i}$ is associated with chiral symmetry breaking in the
unbroken factor $\text{U}(N_{i})$ of the gauge group. The phase
structure of this model is extremely rich and has been much studied in
the literature \cite{ferirr,fer1,fer2,csw1}. The standard analysis
yields two phase invariants: the rank $r$ and the confinement index
$t$. The rank of $|N_{1},k_{1};\ldots;N_{d},k_{d}\rangle$ is the
number of non-zero integers $N_{i}$. Physically it corresponds to the
rank of the low energy gauge group, which is $\u^{r}$ taking into
account the mass gap in the simple non-abelian factors
$\text{SU}(N_{i})$ for $N_{i}\geq 2$. The confinement index $t$ is
defined to be the smallest integer such that the $t^{\text{th}}$
tensor product of the fundamental representation of $\uN$ does not
confine \cite{csw1}. It can be shown that the confinement index in the
vacuum $|N_{1},k_{1};\ldots;N_{d},k_{d}\rangle$ is given by
\be\label{tform} t = N_{1}\wedge\cdots\wedge N_{d}\wedge
(k_{1}-k_{2})\wedge\cdots\wedge (k_{1}-k_{d})\, ,\ee
where $a\wedge b$ represents the greatest common divisor of two
integers $a$ and $b$ \cite{csw1}. It is elementary to show directly on
the solution of the model that $r$ and $t$ are indeed phase
invariants, i.e.\ that they cannot change under analytic continuation
(see for example the discussion in Section 5.2.2 of \cite{ferirr}).
The main result of the present paper will be to construct a new phase
invariant, the Galois group.

The plan of the paper is as follows. In Section \ref{backS}, we review
some background material from \cite{ferirr} that will be useful for
our analysis. In Section \ref{GsymS} we revisit the concept of gauge
symmetry breaking. The usual notion does not make sense at the quantum
level, but we show that a related notion actually does. This leads to
the Galois symmetries. We also briefly explain the relation to Galois
theory of algebraic equations. In Section \ref{GpropS} we analyse some
basic properties of the Galois groups associated with the phases,
showing explicitly that they are phase invariants and that they are
intrinsinc characteristics of the phases. In Section \ref{exS} we
provide a few explicit calculations in simple cases. Finally we
conclude in Section \ref{concS}.

\smallskip

\noindent\emph{Important remark}: the reader is not assumed to be
familiar with Galois theory. We have included in the discussion all
the required notions, which are fairly elementary, from a physicist
point of view. Excellent references on Galois theory are listed in
\cite{Galoisref}.

\smallskip

\noindent\emph{Notation and terminology}: in the following, a
\emph{field} is a ring in which every non-zero element has an inverse.
If $R$ is a ring (which may be a field), we denote by
$R[X_{1},\ldots,X_{n}]$ the polynomial ring in $n$ variables. If $K$
is a field, we denote by $K(X_{1},\ldots,X_{n})$ the field of rational
functions in $n$ variables.

\section{Background material}
\label{backS}

Let us consider an arbitrary $\nn=1$ supersymmetric gauge theory in
four dimensions. Let us denote by $\g$ the set of parameters that
enter in the tree-level superpotential and by $\q$ the set of
instanton factors that are associated with each simple non-abelian
factor of the gauge group. In the case of the $\uN$ theory with one
adjoint and tree-level superpotential $W$ with $W'$ given by
\eqref{Wtree}, one has $\g = (g_{0},\ldots,g_{d})$ and there is a
unique instanton factor given in terms of the dynamically generated
scale $\La$ by $q=\La^{2N}$. The parameters $\g$ and $\q$ are best
viewed as background chiral superfields. The \emph{ring of parameters}
$\pring$ of the theory is defined to be the polynomial
ring\footnote{As explained in great details in \cite{ferirr}, there
are cases where this ring must be enlarged, in particular in theories
that have a zero $\beta$ function. These cases will not enter into the
present paper.}
\be\label{ringparadef} \pring = \mathbb C[\g,\q]\, .\ee

\paragraph{The chiral ring} Any chiral operator in the theory can be
expressed as a finite sum of finite products of a finite set of
generators $\mathcal O_{1},\ldots,\mathcal O_{n}$. In other words, any
chiral operator $\mathcal O$ can be written in the form
\be\label{rhodef}\mathcal O = \rho_{\mathcal O}(\mathcal
O_{1},\ldots,\mathcal O_{n})\, ,\ee
where $\rho_{\mathcal O}\in\pring[X_{1},\ldots,X_{n}]$ is a polynomial
in $n$ variables and coefficients in $\pring$. For example, in the
$\uN$ theory with the adjoint field $\phi$, the generators $\mathcal
O_{i}$ can be taken to be the
\be\label{ukdef} u_{k}=\Tr \phi^{k}\ee
for $1\leq k\leq N$ together with the generalized glueball operators
$v_{k}=-\frac{1}{16\pi^{2}}\Tr W^{\alpha}W_{\alpha} \phi^{k}$ for
$0\leq k\leq N-1$.

An \emph{operator relation} between chiral operators $\mathcal
O^{(1)},\ldots,\mathcal O^{(p)}$ is a relation of the form
\be\label{oprel} P(\mathcal O^{(1)},\ldots,\mathcal O^{(p)}) = 0\, ,\ee
where $P\in\pring[X_{1},\ldots,X_{p}]$, which is valid in all the
vacua of the theory. This definition is unambiguous because, as is
well-known, chiral operators expectation values factorize, for example
$\langle P(\mathcal O^{(1)},\ldots,\mathcal O^{(p)})\rangle =
P(\langle\mathcal O^{(1)}\rangle,\ldots,\langle\mathcal
O^{(p)}\rangle)$. A crucial constraint on the form of an operator
relation is that the coefficients of the polynomial $P$ in
\eqref{oprel} must be in $\pring$, i.e.\ must be polynomials in the
parameters $\g$ and $\q$. As explained in \cite{ferirr}, the full set
of operator relations always follows from a finite number of relations
between a set of generators,
\be\label{relations} P_{i}(\mathcal O_{1},\ldots,\mathcal O_{n}) = 0\,
,\quad P_{i}\in\pring[X_{1},\ldots,X_{n}]\, ,\quad 1\leq i\leq m\, .\ee
A fundamental result derived in \cite{ferirr} is that, in all cases,
the expectation values of the generators (and thus of any other chiral
operator from \eqref{rhodef}) are unambiguously fixed by the relations
\eqref{relations}.

The ring generated by the chiral operators is called the quantum
chiral ring $\ring$. The equations \eqref{rhodef} and
\eqref{relations} show that this ring is a quotient ring given by
\be\label{CR}\ring = \pring[X_{1},\ldots,X_{n}]/\ideal = \mathbb
C[\g,\q,X_{1},\ldots,X_{n}]/\ideal\, ,\ee
where $\ideal$ is the ideal generated by the full set of operator
relations,
\be\label{idealform}\ideal = (P_{1},\ldots,P_{m})\, .\ee
The quantum chiral ring $\ring$ defined above cannot have nilpotent
elements \cite{ferirr}, i.e.\ $\mathcal O^{s}=0$ in $\ring$ for some
$s\in\mathbb N$ implies that $\mathcal O=0$. Equivalently, the ideal
$\ideal$ is radical.

Note that the knowledge of the chiral ring $\ring$ is \emph{equivalent}
to the knowledge of the full solution of the theory in the chiral
sector. Indeed, the ring $\ring$ can be constructed from the knowledge
of the expectation values $\langle\mathcal O\rangle$ for all $\mathcal
O$ and conversely, the full set of expectation values can be found
from the structure of the chiral ring via the relations \eqref{relations}
\cite{ferirr}.

\paragraph{Geometry} The polynomial equations \eqref{relations} define
an affine algebraic variety $\M$ that we shall call the \emph{chiral
variety} of the theory. The chiral ring $\ring$ corresponds to the
ring of regular functions defined on $\M$, also called the coordinate
ring of $\M$,
\be\label{AM} \ring=\mathbb C[\M]\, .\ee
There are thus three \emph{equivalent} ways to present the solution of
the model: a standard way by giving the full set of expectation
values; an algebraic way by giving the ring $\ring$; a geometric way
by giving the variety $\M$.

To make the link between the geometric perspective and the standard
notions of vacua and expectation values, one can consider the
intersection of the space $\M$ with the linear spaces of constant
parameters $\g$ and $\q$. Generically, there is a finite number $v$ of
intersection points, corresponding to a finite number of solutions to
the algebraic equations \eqref{relations} in which the parameters $\g$
and $\q$ are fixed to some particular complex numbers. Each solution
yields the expectation values in a particular vacuum $|i\rangle$ of
the theory. In this picture, the variety $\M$ is thus viewed as a
$v$-fold cover of $\mathbb C^{\delta}$, where $\delta$ is the total
number of parameters $\g$ and $\q$, and the expectation values are
$v$-valued analytic functions of $\g$ and $\q$. In particular, on each
sheet (i.e.\ in each vacuum) there is a semi-classical region
corresponding to small values of $\q$, with a certain pattern of gauge
symmetry breaking and chiral symmetry breaking.

In some special theories, the equations \eqref{relations}, instead of
having a finite number of solutions for fixed $\g$ and $\q$, have a
continuum of solutions corresponding to a moduli space of vacua. The
variety $\M$ may then be viewed as a fibered space with base $\mathbb
C^{\delta}$, the fiber over a point $(\g,\q)$ being the moduli space
of vacua for given parameters $\g$ and $\q$.

\paragraph{Phases} The description of $\M$ as a $v$-fold cover of
$\mathbb C^{\delta}$ is quite arbitrary. Equivalently, the usual
notion of a vacuum is quite arbitrary. A much better defined concept
in the full quantum theory is the notion of \emph{phase}
\cite{ferirr}. The chiral variety $\M$ decomposes into
\emph{irreducible} components according to the phase structure of the
model,
\be\label{Mirrdec}\M=\bigcup_{\varphi}\M_{|\varphi)}\, .\ee
Each irreducible component $\M_{|\varphi)}$ corresponds to a given
phase, denoted by $|\varphi)$, of the theory \cite{ferirr}.

A phase may contain several vacua. A fundamental property is that one
can always interpolate smoothly between two vacua in the same phase by
performing analytic continuations along closed loops in $(\g,\q)$
space \cite{fer1,fer2,csw1,ferirr}. Conversely, vacua in different
phases \emph{cannot} be smoothly connected to each other. Different
irreducible components of $\M$ may intersect, but going from one
component to the other is a non-analytic process associated with a
second order phase transition.

The phase diagram of the theory, or equivalently the decomposition
\eqref{Mirrdec}, can be derived in principle by performing the most
general analytic continuations on the expectation values, finding in
this way which vacua can be smoothly connected to each other and which
cannot. This method is very cumbersome to implement in non-trivial
cases, but fortunately the problem has an equivalent algebraic
formulation that turns out to be much more powerful \cite{ferirr}.
Finding the decomposition \eqref{Mirrdec} is equivalent to decomposing
the ideal $\ideal$ of operator relations into prime ideals,
\be\label{Idealdec}\ideal=\bigcap_{\varphi}\ideal_{|\varphi)}\, .\ee
In practice, this can often be done by factorizing suitable
polynomials into irreducible factors, which yields an elegant and
effective method to compute the phase diagram \cite{ferirr}.

Vacua in a given phase can look very different. For example, they can
have different patterns of gauge symmetry breaking \cite{csw1},
showing most clearly that the gauge group is not a real symmetry and
cannot be used to classify the phases. In the following, our main goal
will be to uncover an underlying symmetry shared by all the vacua in a
given phase.

\paragraph{The field of a phase} In general, the chiral ring $\ring$
defined by \eqref{CR} has zero divisors, i.e.\ non-zero elements
$\mathcal O$ and $\mathcal O'$ such that
\be\label{zerodiv}\mathcal O\mathcal O'=0\, .\ee
This possibility is directly related to the existence of several
phases. Indeed, \eqref{zerodiv} implies that one can split the vacua
into two sets, those for which $\langle\mathcal O\rangle = 0$ but
$\langle\mathcal O'\rangle\not = 0$ and those for which
$\langle\mathcal O'\rangle = 0$ but $\langle\mathcal O\rangle\not =
0$. Clearly one cannot join these two sets of vacua by analytic
continuation. Morevoer, we see that in a given phase there are new
relations, like $\mathcal O=0$ or $\mathcal O'=0$, that are valid in
all the vacua of the phase but not in all the vacua of the theory. The
full set of relations valid in the phase $|\varphi)$ is generated by
the ideal $\ideal_{|\varphi)}$ in the decomposition \eqref{Idealdec}. 
This ideal is \emph{prime}, which means that the chiral ring in the
phase $|\varphi)$,
\be\label{CRphase}\ring_{|\varphi)}=
\pring[X_{1},\ldots,X_{n}]/\ideal_{|\varphi)}\, ,\ee
has no zero divisor. This is equivalent to the irreducibility of
$\M_{|\varphi)}$ in \eqref{Mirrdec}, and we have
\be\label{Mphiring} \ring_{|\varphi)} = \mathbb C[\M_{|\varphi)}]\,
.\ee

Rings that do not have zero divisors are called integral domains. The
simplest example is the ring of integers $\mathbb Z$. Their
fundamental property is that one can consistently consider fractions
of the elements of the ring and build a field of fractions in the same
way as one builds the field of rational numbers $\mathbb Q$ from
$\mathbb Z$. The field of fraction of $\ring_{|\varphi)}$ will be
called the \emph{chiral field in the phase $|\varphi)$} and denoted by
$\field_{|\varphi)}$. It is simply the field of rational functions on
the irreducible variety $\M_{|\varphi)}$,
\be\label{Kphi}\field_{|\varphi)} =
\text{Frac}(\ring_{|\varphi)}) = \mathbb C(\M_{|\varphi)})\,
.\ee
When the theory has a finite number of vacua, this field has an
extremely simple description. There always exists an operator $\mathcal
O_{|\varphi)}$, called a primitive operator in \cite{ferirr} (there
are many primitive operators in a given phase), such that
\be\label{simpleK} \field_{|\varphi)} = \pfield[\mathcal
O_{|\varphi)}] = \pfield[X]/(P_{\mathcal O_{|\varphi)}})\, ,\ee
where 
\be\label{fieldparadef} \pfield = \text{Frac}(\pring)=\mathbb
C(\g,\q)\ee
is the field of parameters of the theory. The result \eqref{simpleK}
is very powerful. It means that in a given phase, \emph{all} the
chiral operators can be expressed as simple polynomials with
coefficients in $\pfield$ of a given primitive operator $\mathcal
O_{|\varphi)}$. All the non-trivial structure is contained in the
single irreducible polynomial equation $P_{\mathcal O_{|\varphi)}}=0$
satisfied by $\mathcal O_{|\varphi)}$ in the phase $|\varphi)$. The
irreducibility of this equation ensures that any rational function of
the chiral operators can always be rewritten as a polynomial in
$\mathcal O_{|\varphi)}$.

\paragraph{What one needs to remember} Chiral operators satisfy a set
of algebraic equations generated by an ideal $\ideal$ that determines
completely the expectation values in all the vacua of the theory.
These algebraic equations define an algebraic variety $\M$, and the
chiral operators are simply the regular functions on this variety.
Phases $|\varphi)$ of the gauge theory correspond to irreducible
components $\M_{|\varphi)}$ of $\M$. On $\M_{|\varphi)}$, chiral
operators satisfy additional relations generated by a prime ideal
$\ideal_{|\varphi)}\supset\ideal$. Thanks to these relations, the
chiral operators in a given phase are actually elements of a field
$\field_{|\varphi)}$, the chiral field in the phase $|\varphi)$, which
is the field of rational functions on $\M_{|\varphi)}$. This fact will
be of utmost importance in the following. When the theory has a finite
number of vacua, the chiral field in a phase is generated by a single
operator and has thus the very simple description \eqref{simpleK}.

\section{Galois symmetries}
\label{GsymS}

We are now going to explain how Galois' ideas allow to circumvent, in
a very subtle and interesting way, the argument showing that the gauge
group cannot be seen in the algebra of local observables. The idea is
to build the gauge variant quantities, as for example the eigenvalues
of the adjoint field $\phi$, from the gauge invariant algebraic
equations these quantities must satisfy. It turns out that the gauge
symmetry can be spontaneously broken by this construction. The
unbroken symmetry, which is a subgroup of the gauge group, is a group
of automorphisms called the Galois group.

\subsection{General considerations}
\label{generaSec}

In the quantum gauge theory, we have direct access to particular
combinations of the matrix elements $\phi_{ij}$, $1\leq i,j\leq N$,
that are gauge invariant,
\be\label{ukphiij} u_{k} = \Tr\phi^{k}=
\phi_{i_{1}i_{2}}\phi_{i_{2}i_{3}}\cdots\phi_{i_{k}i_{1}}\, ,\ee
but not to the matrix $\phi$ itself on which the gauge group acts.
Since our goal is to use the gauge group, we need to reconstruct the
gauge variant objects $\phi_{ij}$. Note that this is an extremely
simple case of the much more general problem of reconstructing open
string degrees of freedom from closed strings. The puzzle is that, to
build the $\phi_{ij}$, we can use only the physical information
contained in the gauge invariants $u_{k}$.

The key to the solution of this problem is to characterize the
$\phi_{ij}$ by the set of algebraic equations \emph{with gauge
invariant coefficients} that they must satisfy. This set of equations
is not difficult to find in our case. Let us introduce the
characteristic polynomial
\be\label{Cpol} C(z) = \det (z-\phi) = z^{N}+\sum_{k=1}^{N}
(-1)^{k}\sigma_{k}z^{N-k}\, ,\ee
where the $\sigma_{k}$s can be expressed in terms of the $u_{k}$s
using the standard Newton's formulas,
\be\label{Newton} \sigma_{1}=u_{1}\, ,\ \sigma_{2} =
\frac{1}{2}\bigl( u_{1}^{2} - u_{2}\bigr)\, ,\ \sigma_{3} =
\frac{1}{6}\bigl(u_{1}^{3}-3u_{1}u_{2}+2u_{3}\bigr)\, ,\ldots\ee
The Cayley-Hamilton theorem implies that the matrix equation
\be\label{Cayley} C(\phi) = \phi^{N}+\sum_{k=1}^{N}
(-1)^{k}\sigma_{k}\phi^{N-k}= 0\ee
must be valid. This yields $N^{2}$ algebraic equations for the
matrix coefficients $\phi_{ij}$. The idea is then to study the
symmetry properties of the algebraic structure that describes the
extension from the set of variables $u_{k}$ to the set of variables
$\phi_{ij}$ governed by the equations \eqref{Cayley}.

In order to simplify the discussion, while keeping the main relevant
features, we are going to focus on the symmetric subgroup
$\SN\subset\uN$ of the gauge group, generated by the permutation
matrices $U_{ij}=\delta_{i\sigma(j)}$ for $\sigma\in\SN$. This means
that instead of decomposing the traces in terms of the $\phi_{ij}$ as 
in \eqref{ukphiij}, we are going to use the simpler decomposition
\be\label{xdef} u_{k} = \Tr \phi^{k} = \sum_{i=1}^{N}x_{i}^{k}\, ,\quad
1\leq k\leq N\, .\ee
The group $\SN$ acts by permuting the $x_{i}$s,
\be\label{SNacts} \sigma :\ x_{i}\longmapsto x_{\sigma(i)}\, .\ee
The $x_{i}$s are the roots of the characteristic polynomial,
\be\label{Cxi} C(x_{i}) = 0\, ,\quad 1\leq i\leq N\, .\ee
This gauge invariant set of equations fully characterize the $x_{i}$s,
in the sense that 
\be\label{Cdecx} C(z) =\prod_{i=1}^{N}(z-x_{i})\ee
immediately implies that
\be\label{defsigma} \sigma_{k} = \sum_{i_{1}<\cdots<i_{k}}
x_{i_{1}}\cdots x_{i_{k}}\, ,\quad 1\leq k\leq N\, ,\ee
and then \eqref{xdef} follows from Newton's formulas. The $x_{i}$s may
be interpreted as being the eigenvalues of the matrix $\phi$, but this
is not necessary and $\phi$ does not need to be diagonalizable. The
important point is that \eqref{xdef} is equivalent to \eqref{Cxi},
from purely algebraic manipulations.

At first sight, the above construction does not seem very promising to
achieve our goal of classifying the phases using the gauge symmetry.
It is true that we can define the $x_{i}$s from the gauge invariant
data, since the equation $C(z)=0$ clearly is gauge invariant, but it
looks like that we really obtain the $x_{i}$s only modulo the action
of the permutation group. It is not obvious to see how some order
parameter, or phase invariant, could emerge from such a construction.

In order to gain some insight into how things might work out, let us
look at the classical theory. The matrix $\phi$ must satisfy the
constraint
\be\label{cleq} W'(\phi) = 0=\prod_{i=k}^{d}(\phi-w_{k})\ee
in this case. The most general solution is labeled as
$|N_{1};\ldots;N_{d}\rangle$, where $N_{k}$ gives the number of
$x_{i}$ equal to $w_{k}$. One says that the gauge group $\uN$ is
broken down to $\text{U}(N_{1})\times\cdots\times\text{U}(N_{d})$ for
this particular solution. The unordered set of integers $\{N_{k}\}$
completely characterizes the phase at the classical (or perturbative)
level. The crucial point here is that all this non-trivial information
is actually contained in the equations \eqref{Cxi}, without the need
to use the matrix $\phi$ explicitly. Indeed, the integers $N_{k}$
simply give the multiplicity of the roots of the gauge invariant
characteristic polynomial $C$ in \eqref{Cpol}. These multiplicities
are coded in the special algebraic relations the gauge invariant
$u_{k}$s (or $\sigma_{k}$s) satisfy. For example, if there is only one
double root ($\uN$ broken down to $\text{U}(2)\times\u^{N-2}$), then
the discriminant of the polynomial must vanish. This constraint on the
coefficients of $C$ characterizes the
$\uN\rightarrow\text{U}(2)\times\u^{N-2}$ pattern of gauge symmetry
breaking, yet it is completely gauge invariant. More generally, for a
given set of integers $\{N_{k}\}$, the coefficients of $C$ will
satisfy a set of constraints generating a certain prime ideal and
corresponding to a certain irreducible algebraic variety. These
constraints are equivalent to the fact that the characteristic
polynomial takes the form
\be\label{Cclass} C(z) = \prod_{k} (z-w_{k})^{N_{k}}\ee
and can be derived by eliminating the $w_{k}$ from the equations
obtained by equating the coefficients of various powers of $z$ in
\eqref{Cclass}.

In the quantum theory, the coefficients of the characteristic
polynomial will still satisfy special algebraic constraints in a given
phase. As reviewed in Section \ref{backS}, these constraints generate
a prime ideal $\ideal_{\vp}$. However, these constraints will no
longer imply that $C$ has multiple roots and the $x_{k}$ will be all
distinct. In other words, the usual notion of gauge symmetry breaking
will be useless in the quantum theory. This is of course not
surprising, since it is known that the set of integers $\{N_{k}\}$ is
not a phase invariant in the full quantum theory. However, Galois
theory teaches us that we still have a well-defined and non-trivial
notion of gauge symmetry breaking that can be used to classify the
phases, as we now explain.

\subsection{The Galois group}
\label{GgroupSec}

According to Section \ref{backS}, in a given phase $\vp$, the gauge
invariant observables belong to a field $\field_{\vp}$, the chiral
field in the phase $\vp$. In particular, the coefficients of the
characteristic polynomial $C$ are elements of $\field_{\vp}$,
\be\label{Cinphase} C\in\field_{\vp}[X]\, .\ee
\subsubsection{The space of eigenvalues}
\label{eigenspace}

As a first step, we need to define precisely the space in which the
roots $x_{i}$ of the characteristic polynomial live. The construction
is based on the following theorem.
\begin{thm}\label{Splitting} Let $Q\in K[X]$ be a polynomial
with coefficients in some field $K$. Then there exists an extension
field $S\supset K$, unique up to field isomorphisms, such that\\
(i) $Q$ factorizes into linear factors in $S$, i.e.\ $Q(z) =
\prod_{k=1}^{N}(z-x_{k})$ with $x_{k}\in S$.\\
(ii) $S$ is generated by the roots of $Q$ over $K$,
$S=K(x_{1},\ldots,x_{N})$.\\
The field $S$ is called the splitting field of the polynomial $Q$.
\end{thm}
We shall denote by $\split_{\vp}$ the splitting field of the
characteristic polynomial $C$ in the phase $\vp$. Note that even
though the roots of $C$ are not gauge invariant, the notion of the
splitting field is a perfectly well-defined and gauge invariant
concept, since it is uniquely determined in terms of the gauge
invariant polynomial $C\in\field_{\vp}[X]$. In physical terms, Th.\
\ref{Splitting} provides a very precise statement about the
construction of gauge-variant data (or ``open string variables''),
which are the roots $x_{i}$, from gauge invariant data (the ``closed
string variables''), which are the completely symmetric polynomials in
the roots. The splitting field is the space in which the gauge-variant
data lives.

\subsubsection{Gauge symmetry breaking \`a la Galois}
\label{breakGalois}

The construction of $\split_{\vp}$ is perfectly $\SN$-symmetric, yet,
as Galois realized, this symmetry can be spontaneously broken. This
yields a new version of the concept of gauge symmetry breaking that
makes sense in the quantum theory and that, as we shall explain in
details in the following, is a phase invariant.

Galois' notion of symmetry breaking can be introduced as follows.
Consider a polynomial $Q\in K[X]$, where $K$ is a field, and let
$S=K(x_{1},\ldots,x_{N})$ be its splitting field. Let $\mathscr O$ be
an arbitrary element of $S$. $\mathscr O$ is a rational function of
the roots with coefficients in $K$, $\mathscr O=\mathscr
O(x_{1},\ldots,x_{N})$. Let us ask the following question: what
conditions do we need to impose on $\mathscr O$ to ensure that
$\mathscr O\in K$? Clearly, a \emph{sufficient} condition is to impose
full symmetry of $\mathscr O$ under arbitrary permutations of the
roots. This condition is also necessary if $Q$ is a generic polynomial
(i.e.\ a polynomial for which the coefficients are algebraically
independent over $K$). However, this condition is \emph{not} necessary
in general. For example, a given $Q$ might have some of its roots in
$K$, in which case \emph{any} rational function of these roots is of
course in $K$. One of the main result in Galois theory is to show that
the necessary and sufficient condition for $\mathscr O\in S$ to be in 
$K$ is governed by a symmetry principle.

More precisely, the symmetric subgroup $\SN$ of the
gauge group acts on $\mathscr O\in S$ as
\be\label{Gacts}\sigma\cdot \mathscr O(x_{1},\ldots,x_{N}) =
\mathscr O(x_{\sigma(1)},\ldots,x_{\sigma(N)})\, .\ee
We then have the following theorem.
\begin{thm}\label{fundthm} Let $K$ be a field of characteristic zero
(i.e.\ that contains $\mathbb Q$ as a subfield), $Q\in K[X]$ and $S$
be its splitting field. There always exists a subgroup $G$ of the
permutation group $\SN$, called the Galois group of the polynomial
$Q$, such that, for any $\mathscr O\in S$,
\be\label{fundequiv} \mathscr O\in K \Longleftrightarrow \sigma\cdot
\mathscr O=\mathscr O\ \text{for all}\ \sigma\in G\, .\ee
\end{thm}

In our case, $K=\field_{\vp}$ is automatically of characteristic zero
(since it contains $\mathbb C$) and the polynomial we consider is the
characteristic polynomial.

\smallskip

\noindent\textbf{Definition.} The \emph{Galois group $G_{|\varphi)}$
of the phase $\vp$} is the Galois group of the characteristic
polynomial $C\in\fieldp[X]$.

\smallskip

Theorem \ref{fundthm} means that $G_{\vp}$-invariant operators are
physical. This may be confusing, since being $G_{\vp}$-invariant is in
general a \emph{weaker} statement than being $\SN$-invariant, which is
the standard condition for gauge invariance. However, it is important
to realize that, even when $G_{\vp}$ is a strict subgroup of $\SN$,
the set of observables is not larger than usual and is still given by
the completely symmetric functions in the roots. The theorem simply
ensures that any $G_{\vp}$-symmetric operator can be expressed
unambiguously in terms of $\SN$-symmetric operators, or in other words
that the $G_{\vp}$-symmetric operators are completely fixed in terms
of gauge invariant data only.

Very interestingly, the fact that the Galois group can be a strict
subgroup of $\SN$ is just an example of the familiar concept of
symmetry breaking. Symmetry breaking occurs when then equations of the
problem under consideration are symmetric, but the solution is not. In
the case of a polynomial $Q\in K[X]$, the problem is the construction
of the splitting field $S$ in terms of $Q$ and $K$. The equations that
determine $S$ are $\SN$-symmetric, but the structure of $S$ may not be
$\SN$-symmetric. In mathematical terms, this means that the group of
automorphisms of $S$ may be smaller than $\SN$. 

The precise statements are as follows. Let $\text{Gal}(S|K)$ be the
group of $K$-auto\-mor\-phisms of $S$, i.e.\ automorphisms of $S$ that act
trivially on $K$. The polynomial $Q\in K[X]$ is clearly invariant
under any element of $\text{Gal}(S|K)$ and thus a $K$-automorphism
acts by permuting the eigenvalues $x_{i}$. Moreover, a
$K$-automorphism is completely determined by the permutation it
induces on the roots, since $S=K(x_{1},\ldots,x_{N})$. One then has
the following theorem.
\begin{thm}\label{autothm} The group of automorphisms of $S$ that 
let $K$ fixed coincides with the Galois group of the polynomial $Q$ as
defined in Th.\ \ref{fundthm},
\be\label{GQGal} G = {\rm Gal}(S|K)\, .\ee
\end{thm}
Note that in modern treatments of Galois theory, one usually starts
from the group $\text{Gal}(S|K)$. The non-trivial part of Th.\
\ref{autothm} is then to show that the field of
$\text{Gal}(S|K)$-invariant rational functions of the roots is $K$ and
not a larger field. This doesn't work for an arbitrary field
extension, but it works in characteristic zero for the so-called
normal extensions. The basic example of a normal extension is
precisely the field extension corresponding to the splitting field of
a polynomial.

Let us repeat once more the fundamental property of the Galois group,
in physical terms. If $\sigma\cdot {}$ denotes the action of an
element of the gauge group, here the symmetric group $\SN$, the
observables are usually constrained to satisfy
\be\label{kinaction} \sigma\cdot\mathscr O = \mathscr O\quad \text{for
any }\sigma\in \SN\, .\ee
However, the ``kinematical'' constraint \eqref{kinaction} is not really
justified. The physical requirement is rather that the expectation
values should be gauge invariant,
\be\label{dynaction} \bigl\langle \sigma\cdot\mathscr O\bigr\rangle =
\langle\mathscr O\rangle\quad \text{for any }\sigma\in \SN\, .\ee
Of course, \eqref{kinaction} implies \eqref{dynaction}, but the
converse is not necessarily true. Due to dynamical constraints, it can
happen that kinematical constraints less stringent than
\eqref{kinaction} automatically imply the full gauge invariance
\eqref{dynaction} of the expectation values. From our previous
discussion, we know that in a given phase $\vp$, the kinematical
constraints that ensure the validity of \eqref{dynaction} take the
form
\be\label{kinactionG} \sigma\cdot\mathscr O = \mathscr O\quad
\text{for any }\sigma\in G_{\vp}\, ,\ee
where $G_{\vp}$ is the Galois group of the phase $\vp$. Indeed, the
condition \eqref{kinactionG} ensures that $\mathscr O\in\fieldp$,
i.e.\ $\langle\mathscr O\rangle=\langle\mathcal O\rangle$ in the phase
$\vp$, where $\mathcal O$ is a standard gauge invariant (i.e.\
completely symmetric) operator.

\subsubsection{Relation with the solution of algebraic equations by
radicals}

In order to assuage a natural curiosity the reader may have, let us
very briefly discuss the link with the theory of algebraic equations
that are solvable by radicals, which is the original and most famous
application of Galois theory.

An algebraic equation is said to be solvable by radicals when its
roots can be expressed in terms of the coefficients by a formula that
involves only ordinary additions, multiplications and extraction of
$p^{\text{th}}$ roots, for any $p$. It is possible to show that if one
of the root of an irreducible polynomial has this property, all the
other roots will also have it.

If an algebraic equation is solvable by radicals, then its Galois
group $G$ cannot be an arbitrary group. If $G^{(1)}$ is the commutator
subgroup of $G$ (the subgroup generated by all the commutators of
pairs of elements of $G$), $G^{(2)}$ the commutator subgroup of
$G^{(1)}$ and so on, then there must exist a $k\geq 1$ such that
$G^{(k)}$ is trivial. One then says that the group $G$ is solvable.

Thus, if the Galois group of a particular equation is not solvable, we
can immediately deduce that the equation cannot be solved by radicals.
For example, the generic equation of degree $N$ has Galois group
$G=\SN$ which is not solvable for $N\geq 5$ ($\SN^{(k)}=\text{A}_{N}$
the group of even permutations for all $k\geq 1$, because
$\text{A}_{N}$ is simple for $N\geq 5$). Thus a general formula
involving only radicals for the roots of an arbitrary polynomial of
degree greater than five does not exist.

Galois' idea is very similar to Landau's idea for classifying the
phases. Laudau argues that two vacua in the same phase must have the
same symmetry group, whereas Galois argues that all equations solvable
by radicals must have solvable Galois groups. In the case of gauge
theories, we are going to derive shortly that two phases must be
distinct if they have distinct Galois symmetries.

Galois theory of algebraic equations is even more powerful, because it
gives a complete solution to the problem (whereas we do not claim that
the Galois symmetries of the phases yield a complete classification):
if the Galois group of an algebraic equation is solvable, then the
equation is solvable by radicals. The existence of equations of degree
greater than five solvable by radicals of course does not contradict
the result for the generic equation, because the formulas for the
roots will be equation-dependent.

\subsubsection{Galois groups and monodromy groups}
\label{monoSec}

We are now going to use the fact that, in the context of gauge
theories, the base field $\fieldp$ is the field of rational functions
of the irreducible variety $\M_{\vp}$, see \eqref{Kphi}. This yields a
very interesting characterization of the Galois group, that we shall
use later in explicit computations.

The general problem is as follows. We consider a degree $N$ polynomial
$Q\in K[X]$, where $K=\mathbb C(V)$ is the field of rational functions
on some irreducible algebraic variety $V$. Let $\mathcal C$ be a
closed contour on $V$. Since the coefficients of $Q$ are single-valued
on $V$, $Q$ is mapped onto itself if we perform an analytic
continuation along $\mathcal C$. This shows that any given root of $Q$
must be mapped onto another root under the analytic continuation.
Thus, to any closed contour $\mathcal C$ in $V$, we can associate a
permutation $\sigma_{\mathcal C}$ of the roots of $Q$ obtained by
performing the analytic continuation along $\mathcal C$. The subgroup
of $\SN$ generated by the $\sigma_{\mathcal C}$ for all closed
contours $\mathcal C\subset V$ is called the \emph{monodromy group} of
the polynomial $Q\in\mathbb C(V)[X]$.
\begin{thm}\label{monodGalois} The monodromy group of $Q\in\mathbb
C(V)[X]$ coincides with the Galois group of $Q$.
\end{thm}
\noindent Let $G_{Q}$ denotes the Galois group as usual, and $\tilde
G_{Q}$ denotes the monodromy group. Let the $x_{i}$s be the roots of
$Q$ and $S_{Q}=\mathbb C(V)(x_{1},\ldots,x_{N})$ denotes the splitting
field. To any $\sigma_{\mathcal C}\in\tilde G_{Q}$ we can associate an
automorphism of $S_{Q}$ that let $\mathbb C(V)$ fixed by defining
$\sigma_{\mathcal C}\cdot f(x_{1},\ldots,x_{N}) =
f(x_{\sigma_{\mathcal C}(1)},\ldots,x_{\sigma_{\mathcal C}(N)})$ if
$f\in S_{Q}$. This is well-defined because if $f=a/b$, $a$ and $b\not
=0$ being polynomials, then $\sigma_{\mathcal C}\cdot f =
\sigma_{\mathcal C}\cdot a/\sigma_{\mathcal C}\cdot b$ and clearly
$b\not = 0\Rightarrow \sigma_{\mathcal C}\cdot b\not = 0$. This shows
that $\tilde G_{Q}\subset G_{Q}$. Conversely, consider $f\in S_{Q}$
and assume that $\sigma_{\mathcal C}\cdot f = f$ for all
$\sigma_{\mathcal C}\in\tilde G_{Q}$. This implies that $f$ is a
well-defined meromorphic function on $V$. Moreover, $f$ is
automatically algebraic over $\mathbb C(V)$, i.e.\ $f$ satisfies an
irreducible polynomial equation with coefficients in $\mathbb C(V)$.
Indeed, this is true for any element of $S_{Q}$, because $S_{Q}$ is a
finite extension of $\mathbb C(V)$, of degree $|G_{Q}|\leq N!$ (this
means that the degree of the equation satisfied by $f$ is less than
$N!$). Now, since $f$ is single-valued on $V$, the irreducible
equation that it satisfies must necessarily be of degree one, showing
that $f\in\mathbb C(V)$. From Th.\ \ref{fundthm} we thus get
$G_{Q}\subset\tilde G_{Q}$ and finally $G_{Q}=\tilde G_{Q}$.

\section{Properties of the Galois group of a phase}
\label{GpropS}

In this Section, we discuss two important properties of the Galois
group of a phase. The first property is that that the Galois group can
be computed from the small $q$ expansion of the correlators in any
vacuum. The result is independent of the choice of vacuum in the phase
$\vp$. This makes explicit the fact that the Galois group is a phase
invariant. The second property of the Galois group that we discuss is
its intrinsic nature. We make precise the fact that, unlike the field
$\fieldp$, $G_{\vp}$ does not depend on the particular realization of
the phase in a given model, but only on the phase itself.

\subsection{The Galois group as a phase invariant}
\label{phaseinvSec}

We have defined the Galois group of a phase in a global,
vacuum-independent way in Section \ref{GgroupSec}. We are now going to
exhibit a construction that makes clear that the Galois group can
actually be computed from the knowledge of the solution in an
arbitrary small neighbourhood of the classical limit $q=0$ in any
vacuum belonging to the phase.

Let us thus pick a vacuum $|i\rangle$ in a phase $|\varphi)$. As
explained in Section \ref{backS}, this amounts to choosing a
particular value $\langle i|\mathcal O|i\rangle$ for the multi-valued
analytic functions $\langle \mathcal O\rangle$. In a given phase
$\vp$, we can focus on a primitive operator $\mathcal O_{\vp}$, since
all the other operators are polynomials in $\mathcal O_{\vp}$. The
operator $\mathcal O_{\vp}$ satisfies an irreducible polynomial
equation $P_{\mathcal O_{\vp}}=0$ with coefficients in
$\pfield=\mathbb C(\g,q)$, and each root of this equation corresponds
to a particular vacuum in the phase,
\be\label{POPeq}P_{\mathcal O_{\vp}}(z) =
\prod_{|j\rangle\in\vp}\bigl( z-\langle j|\mathcal
O_{\vp}|j\rangle\bigr)\, .\ee
We can consider the field generated by any of the roots $\langle
i|\mathcal O_{\vp}|i\rangle$ over $\pfield$,
\be\label{fieldvac}\mathsf K_{|i\rangle} = \pfield \bigl(\langle
i|\mathcal O_{\vp}|i\rangle\bigr)\, . \ee
This is a field because $P_{\mathcal O_{\vp}}$ is irreducible. The
expectation value of the characteristic polynomial \eqref{Cpol} in the
vacuum $|i\rangle$ has its coefficients in $\mathsf K_{|i\rangle}$,
$\langle i|C|i\rangle\in\mathsf K_{|i\rangle}[X]$. If $S_{\langle
i|C|i\rangle}$ is the splitting field of $\langle i|C|i\rangle$, we
can define the Galois group in the vacuum $|i\rangle$ to be
\be\label{Gvac} G_{|i\rangle} =
\text{Gal}(S_{\langle i|C|i\rangle}|K_{|i\rangle})\, .\ee

The expectation value $\langle i|\mathcal O_{\vp}|i\rangle$, as any
other chiral operator expectation values in the vacuum $|i\rangle$,
can be Puiseux-expanded at small $q$. The Puiseux expansion is a
convergent expansion in terms of some fractional power of $q$. Of
course this expansion strongly depends on the choice of vacuum. For
example, it can be shown easily that the expansion parameter in a
vacuum characterized by a pattern of gauge symmetry breaking
$\uN\rightarrow\text{U}(N_{1})\times\cdots\times\text{U}(N_{r})$,
where the $N_{i}$s are all non-zero, is $q^{N_{1}\wedge\cdots\wedge
N_{r}/(N_{1}\cdots N_{r})}$ where $N_{1}\wedge\cdots\wedge N_{r}$ is
the greatest common divisor of the $N_{i}$s. The Puiseux expansions
give the local data associated with a vacuum. Clearly, $\langle
i|\mathcal O_{\vp}|i\rangle$ and thus $\mathsf K_{|i\rangle}$ and
$G_{|i\rangle}$ are determined in terms of this local data only.

We can now state the basic result.
\begin{thm}\label{vacGalthm} The Galois groups $G_{|i\rangle}$ are
phase invariants, i.e.\ if $|i\rangle$ and $|j\rangle$ belong to the
same phase $\vp$, then $G_{|i\rangle}=G_{|j\rangle}=G_{\vp}$.
\end{thm}
\noindent This theorem relies on two standard results in Galois
theory,
\begin{lem} i) Let $P\in K[X]$ be an irreducible polynomial, $a$ and
$b$ two roots of $P$. Then there exists a field isomorphism
$f:K(a)\rightarrow K(b)$, $f(a)=b$, that let $K$ fixed. Moreover,
$K(a)$ and $K(b)$ are both isomorphic to $K[X]/(P)$.\\
ii) Let $Q\in K(a)[X]$ and $f(Q)\in K(b)[X]$ its image under $f$. Let
$S_{Q}$ be the splitting field of $Q$ over $K(a)$ and $S_{f(Q)}$ the
splitting field of $f(Q)$ over $K(b)$. Then the field isomorphism $f$
can be extended into a field isomorphism $g: S_{Q}\rightarrow
S_{f(Q)}$.
\end{lem}
\noindent The part i) of the lemma states that, from the algebraic
point of view, all the roots of an \emph{irreducible} polynomial are
indistinguishable. Part ii) of the lemma is a version of the unicity
theorem for the splitting field. In our case, we use the Lemma to
construct various field isomorphisms $f$, $g$, $F$ and $G$ as in the
following diagram. The vertical arrows represent the canonical
inclusions.
\be\begin{CD} S_{\langle i|C|i\rangle} @> g>>
S_{\langle j|C|j\rangle} @> G>> \mathsf S_{|\varphi)}\\
@AAA @AAA @AAA \\
\mathsf K_{|i\rangle} @>f>> \mathsf K_{|j\rangle} @>F>> \fieldp\\
@AAA @AAA @AAA \\
\pfield @>\text{id}>>\pfield @>\text{id}>>\pfield
\end{CD}\ee
Since the Galois groups are groups of field automorphisms, the
isomorphisms between the groups follow immediately. For example, the
isomorphism between $G_{|i\rangle}$ and $G_{|j\rangle}$ is given by
$\sigma\mapsto g\sigma g^{-1}$ and the isomorphism between
$G_{|j\rangle}$ and $G_{\vp}$ is given by $\sigma\mapsto G\sigma
G^{-1}$. This shows explicitly that the Galois group is a phase
invariant.

\subsection{Intrinsic nature of the Galois group}
\label{intrinSec}
\subsubsection{Generalities}

The chiral field $\field_{\vp}$ in a phase $\vp$ does not characterize
the phase $\vp$. In particular, the number of vacua in a phase, which
is the degree of $\field_{\vp}$ over $\pfield$ \cite{ferirr}, can vary
depending on the way the phase is realized. 

For example, consider the case where the derivative of the tree-level
superpotential \eqref{Wtree} is given by $m z$. Then the theory
has $N$ distinct vacua with unbroken gauge group
$|N,k\rangle=|k\rangle$, $0\leq k\leq N-1$, all belonging to the same
confining phase $|C)$. By sending $m$ to infinity, one obtains the
pure $\nn=1$ gauge theory, which is thus also in the phase $|C)$.

More generally, consider the case of an arbitrary tree-level
superpotential \eqref{Wtree}. The $Nd$ vacua of rank one can be
labeled as $|i,k\rangle = |N_{1},k_{1};\ldots;N_{d},k_{d}\rangle$ with
$N_{j}=N\delta_{ij}$ and $k_{i}=k$. It is not difficult to check that
all these vacua belong to the same phase $|\tilde C)$. Clearly, the
number of vacua in $|C)$ and $|\tilde C)$ are different ($N$ and $Nd$
respectively), and the fields $\field_{|C)}$ and $\field_{|\tilde C)}$
are different. However, we would like to think about $|C)$ and
$|\tilde C)$ as describing the same physical phase, since the vacua in
the phase $|C)$ can be obtained by taking an appropriate limit from
the vacua of the phase $|\tilde C)$. For example, the vacuum
$|i,k\rangle\in |\tilde C)$ goes to $|k\rangle\in |C)$ if the
parameters in the tree level superpotential are set such that
$W(z)=\frac{m}{2}(x-w_{i})^{2}$.

The aim of the present subsection is to show that, unlike the chiral
field, the Galois symmetry is an intrinsic property of the phase. In
particular, in the example described above, one has
$G_{|C)}=G_{|\tilde C)}$. Mathematically, we have to study how the
Galois group depends on the base field. We could use a purely
algebraic route based on the so-called theorem on natural
irrationalities (see for example the first reference in
\cite{Galoisref}) but we prefer to present a more analytic approach
based on Theorem \ref{monodGalois}.

\subsubsection{Intrinsic nature of the Galois group}
\label{intrSec}

As we have mentioned in the Introduction section, the rank of the
vacua is a phase invariant and the chiral variety \eqref{Mirrdec}
decomposes accordingly,
\be\label{Mrankdec}\M=\bigcup_{r=1}^{N}\M_{r}\, .\ee
The varieties $\M_{r}$ are described by the following factorization
conditions involving the characteristic polynomial \eqref{Cpol} and
the derivative of the tree-level superpotential \eqref{Wtree} (see
\cite{ferirr} for details and an extensive list of references)
\begin{align}\label{eqI} C(z)^{2} - 4 q &= M_{N-r}(z)^{2}Y_{2r}(z)\, ,\\
\label{eqII} W'(z)^{2} - D_{d-1}(z) &= N_{d-r}(z)^{2}Y_{2r}(z)\, .
\end{align}
In the above equations, $M_{N-r}$, $Y_{2r}$ and $N_{d-r}$ are
polynomials of degrees $N-r$, $2r$ and $d-r$ respectively, whereas
$D_{d-1}$ is a polynomial of degree at most $d-1$. Of course, vacua of
rank $r$ exist only if $d\geq r$.

The equation \eqref{eqI} is the standard factorization condition on
the Seiberg-Witten curve. It simply states that $C^{2}-4q$ has $N-r$
double roots. This condition is equivalent to a set of algebraic
equations satisfied by the coefficients of $C$ over $\mathbb C[q]$.
These equations can be obtained for example by eliminating the
coefficients of $M_{N-r}$ and $Y_{2r}$ from the constraints obtained
by matching the powers of $z$ in \eqref{eqI}. Explicit examples of the
resulting equations are given in Section \ref{exS}. We denote by
$\mathscr N_{r}$ the corresponding $r+1$-dimensional algebraic
variety. Standard local coordinates on $\mathscr N_{r}$ are given by
$q,u_{1},\ldots,u_{r}$. From the analysis in \cite{ferirr}, it is
straightforward to check that for given $q,u_{1},\ldots,u_{r}$ there
are
\be\label{vhatr}\hat v_{r}(N) =
\binom{N+r-1}{2r-1}=\sum_{\sum_{k=1}^{r}N_{k}=N}N_{1}\cdots N_{r}\ee
points on $\mathscr N_{r}$. This is also the minimal number of vacua
in any realization of the phases of rank $r$.

The defining equations of $\M_{r}$ are obtained by combining equations
\eqref{eqI} and \eqref{eqII} and eliminating all variables except the
$u_{k}$s (or equivalently the coefficients of $C$), $q$ and the
$g_{k}$s. It is a $d+2$ dimensional variety with standard local
coordinates $(q,g_{0},\ldots,g_{d})=(q,\g)$. It decomposes in terms of
irreducible components according to
\be\label{Mrdec} \M_{r} = \bigcup_{|r,\varphi)}\M_{|r,\varphi)}\, ,\ee
where we denote by $|r,\varphi)$ the phases at rank $r$.

It is useful to introduce the $r$-dimensional subvariety $\hat\M_{r}$
of $\M_{r}$ corresponding to $g_{k}=0$ for $r+1\leq k\leq d$ and
$g_{r}=1$. Using the results of \cite{ferirr}, it is straightforward
to show that the decomposition in irreducible components of $\M_{r}$
and $\hat\M_{r}$ are in one-to-one correspondence. The idea is that
the non-trivial interpolations between the vacua in $\M_{|r,\varphi)}$
all follow from the non-trivial interpolations between the vacua that
are also in $\hat\M_{|r,\varphi)}$ together with additional
interpolations that can all be described in the semi-classical regime.
One thus has
\be\label{Mrhatdec}\hat\M_{r} =
\bigcup_{|r,\varphi)}\hat\M_{|r,\varphi)}\, ,\ee
where $\hat\M_{|r,\varphi)}$ is the subvariety of $\M_{|r,\varphi)}$
obtained by setting $g_{r+1}=\ldots=g_{d}=0$.

Let us now consider the projection map $\pi: \M_{r}\rightarrow\mathscr
N_{r}$, $\pi(q,\g,u_{1},\ldots,u_{N})= (q,u_{1},\ldots,u_{N})$. $\pi$
is well-defined since the equations on $q,u_{1},\ldots,u_{N}$ that
define $\mathscr N_{r}$ are automatically satisfied on $\M_{r}$.
Moreover, we have the following fundamental property of $\pi$.
\begin{lem}\label{restLem} The restriction of the projection map
$\hat\pi=\pi_{|\hat\M_{r}}:\ \hat\M_{r}\rightarrow\mathscr N_{r}$ is
one-to-one.
\end{lem}
To prove this lemma, let us fix $q,u_{1},\ldots,u_{N}$ in \eqref{eqI}.
This fixes the polynomials $M_{N-r}$ and $Y_{2r}$ in a unique way. If
we assume that $g_{r+1}=\ldots=g_{d}=0$, then \eqref{eqII} implies
that $N_{d-r}=g_{r}^{2}=1$ and that $W'$ is uniquely determined to be
the polynomial part in the large $z$ expansion of $\sqrt{Y_{2r}}$. The
polynomial $D_{d-1} = W'^{2}-Y_{2r}$ has then a degree bounded by
$r-1\leq d-1$ as it should.

The algebraic varieties $\hat\M_{r}$ and $\mathscr N_{r}$ are thus
isomorphic and in particular $\mathscr N_{r}$ decomposes into
irreducible components as
\be\label{Nrdec}\mathscr N_{r}=\bigcup_{|r,\varphi)}\mathscr
N_{|r,\varphi)}\, ,\ee
where $\mathscr N_{|r,\varphi)} = \hat\pi(\M_{|r,\varphi)})$. One then 
has the
\begin{thm}\label{intrinsicthm} (Intrinsic nature of the Galois group)
The Galois group of the phase $|r,\varphi)$, which is the Galois group
of the characteristic polynomial \eqref{Cpol} viewed as a polynomial
with coefficients in the field $\mathbb C(\M_{|r,\varphi)})$, is the
same as the Galois group of the characteristic polynomial \eqref{Cpol}
viewed as a polynomial with coefficients in the field $\mathbb
C(\mathscr N_{|r,\varphi)})$.
\end{thm}
\noindent This theorem not only demonstrates the intrinsic nature of
the Galois group of a phase but also shows that the computation of the
group can be done by considering the varieties $\mathscr
N_{|r,\varphi)}$ only. This will be used in the next Section.

Let us prove the theorem by using the characterization of the Galois
group given in Th.\ \ref{monodGalois}. We consider the characteristic
polynomial $C\in\mathbb C(\M_{|r,\varphi)})[X]$ with Galois group
$G_{|r,\varphi)}$ and $\pi(C)\in\mathbb C(\mathscr
N_{|r,\varphi)})[X]$ with Galois group $\tilde G_{|r,\varphi)}$. Let
$\mathcal L$ be a closed contour in $\M_{|r,\varphi)}$, with
associated element $\sigma_{\mathcal L}\in G_{|r,\varphi)}$
corresponding to the permutation of the roots of $C$ obtained by
performing the analytic continuation of $C$ along $\mathcal L$. The
analytic continuation of $\pi(C)$ along the closed contour
$\pi(\mathcal L)\subset\mathscr N_{|r,\varphi)}$ will obviously yield
the same permutation of the roots. Thus $G_{|r,\varphi)}\subset\tilde
G_{|r,\varphi)}$. Conversely, let $\tilde{\mathcal L}$ be a closed
contour in $\mathscr N_{|r,\varphi)}$, with associated element
$\tilde\sigma_{\tilde{\mathcal L}}$. The Lem.\ \ref{restLem} ensures
that there exists a \emph{closed} contour $\mathcal L$ in
$\M_{|r,\varphi)}$ such that $\pi(\mathcal L) = \tilde{\mathcal L}$:
one can choose $\mathcal L = \hat\pi^{-1}(\tilde{\mathcal L})$. This
shows that $\tilde G_{|r,\varphi)}\subset G_{|r,\varphi)}$ and we can
conclude.

\section{A few simple examples of Galois groups}
\label{exS}

Computing Galois groups is, in general, rather subtle. It may be
possible to find the Galois groups for all the phases of the model
\eqref{Wtree}, but we shall be more modest here and limit ourselves to
explicit calculations in a few very simple cases. Our goal is to make
the discussions of the previous Sections as concrete as possible. We
start in \ref{51s} with the case of the $\text{U}(2)$ and
$\text{U}(3)$ theories, which is very elementary. In \ref{52s}, we
present a few general features of the Galois groups valid in $\uN$
theories for any $N$. Finally, in \ref{53s}, we compute the Galois
groups for all the phases of the $\text{U}(4)$ model.

The phases of the model have been studied extensively in \cite{ferirr}
and we shall use the results of this paper as well as of the earlier
references \cite{fer1,csw1,fer2} in the following.

We want to compute the Galois group of the characteristic polynomial
\eqref{Cpol}, where the coefficients satisfy a suitable set of
constraints correponding to the phase under study. At the expense of
shifting the variable $z\rightarrow z+\sigma_{1}/N$ in \eqref{Cpol},
we can always set $\sigma_{1}=0$. Obviously, such a shift of the
indeterminate by an element of the base field doesn't change the
Galois group. In other words, the Galois groups for the $\uN$ or
$\text{SU}(N)$ theories are the same. With $\sigma_{1}=0$, the
irreducible varieties $\mathscr N_{|r,\varphi)}$ of Section
\ref{intrSec} are of dimension $r$ instead of $r+1$.

\subsection{The Galois groups for $\text{U}(2)$ and $\text{U}(3)$}
\label{51s}

In the case of $\text{U}(2)$, the characteristic polynomial
\eqref{Cpol} is
\be\label{C2} C(z) = z^{2} + \sigma_{2}\, .\ee
At rank $r=2$, we have the usual Coulomb phase $|2)$ with unbroken
gauge group $\u^{2}$. The coefficient of $C$ in this phase in not
constrained. The Galois group is thus the same as for the generic
polynomial, $G_{|2)}=\text{S}_{2}=\mathbb Z_{2}$. At rank $r=1$ we
have the confining phase $|1)$ with unbroken gauge group. This is
characterized by the condition $\sigma_{2}^{2}=4q$. Since $q$ is
arbitrary, we are again in the case of a generic polynomial and thus
$G_{|1)}=\text{S}_{2}=\mathbb Z_{2}$.

The above results are extremely simple to interpret. For example, in
the case of the Coulomb phase, the roots of $C$ are given by
$x_{1}=i\sqrt{\sigma_{2}}$ and $x_{2}=-i\sqrt{\sigma_{2}}$. The only
polynomials in the roots that can be expressed in terms of a
polynomial of $\sigma_{2}$ are the symmetric polynomials in $x_{1}$
and $x_{2}$. Thus the Galois group is $\text{S}_{2}$. Another way to
understand the result is to note that the irreducible variety
$\mathscr N_{|2)}$ is just $\mathbb C^{2}$, with coordinates
$\sigma_{2}$ and $q$. The roots $x_{1}$ and $x_{2}$ are exchanged by
performing an analytic continuation along a closed contour that
circles around $\sigma_{2}=0$, $\sigma_{2}\rightarrow
e^{2i\pi}\sigma_{2}$ and this transposition generates the Galois group
$\text{S}_{2}$.

In the case of $\text{U}(3)$,
\be\label{C3} C(z) = z^{3}+\sigma_{2} z - \sigma_{3}\, .\ee
At rank $r=3$, we have the Coulomb phase $|3)$ with a generic
characteristic polynomial and $G_{|3)}=\text{S}_{3}$. At rank $r=2$,
there is a unique phase $|2)$ with irreducible variety $\mathscr
N_{|2)}$ given by the equation
\be\label{N3r2eq} \bigl(27\sigma_{3}^{2}+4\sigma_{2}^{3}\bigr)^{2} -
216q\bigl(27\sigma_{3}^{2}-4\sigma_{2}^{3}\bigr) + 11664 q^{2} = 0\, . \ee
This equation can be viewed as a constraint on $q$ for given
$\sigma_{2}$ and $\sigma_{3}$ but $\sigma_{2}$ and $\sigma_{3}$ can be
choosen freely. As a consequence, $C$ is generic and
$G_{|2)}=\text{S}_{3}$. 

The first interesting case, for which the $\text{S}_{3}$ symmetry is
spontaneously broken, corresponds to the confining phase $|1)$ at rank
$r=1$. The irreducible variety $\mathscr N_{|1)}$ is given by the
equations that ensure that $C^{2}-4q$ has two double roots. This
implies that the discriminant of the polynomials $C-2q^{1/2}$ and
$C+2q^{1/2}$ both vanish, which yields, after some simple algebraic
manipulations,
\be\label{N3r1eq} \sigma_{3}=0\, ,\quad \sigma_{2}^{3}+27 q = 0\, .
\ee
The polynomial $C$ is thus reducible over $\mathbb C(\mathscr
N_{|1)})$, $C(z) = z(z^{2}+\sigma_{2})$. Since $\sigma_{2}$ can be
arbitrary, the Galois group is $G_{|1)}=\mathbb Z_{2}$, exchanging the
two roots of $z^{2}+\sigma_{2}=0$ and letting the third root fixed.

\subsection{Galois groups for $\uN$}
\label{52s}

We can make a few simple general statements about the Galois groups of
some phases for arbitrary $N$.

\subsubsection{The phase at rank $N$}

At rank $r=N$ there is only one phase, the Coulomb phase with unbroken
gauge group $\u^{N}$. The irreducible variety $\mathscr N_{|N)}$ is
simply $\mathbb C^{N}$ in this case and the characteristic polynomial 
is a generic polynomial of degree $N$. The Galois group is thus
automatically the full permutation group,
\be\label{GalCoulomb} G_{|N)} = \SN\, .\ee
\subsubsection{The phase at rank $N-1$}

There is only one phase at rank $N-1$, in which all the vacua have a
$\u^{N-2}\times\text{U}(2)$ unbroken gauge group \cite{ferirr}. The
irreducible variety $\mathscr N_{|N-1)}$ is a hypersurface in $\mathbb
C^{N}$ given by the vanishing of the discriminant of $C^{2}-4q$ (this 
discriminant is automatically an irreducible polynomial from our
previous discussions). This equation can be seen as a contraint on $q$
for given $u_{k}$s, and thus the characteristic polynomial is generic,
which implies that
\be\label{GalNmone} G_{|N-1)} = \SN\, .\ee
\subsubsection{The phase at rank one}

The solution for the unique phase at rank $r=1$, corresponding to
$C^{2}-4q$ having $N-1$ double roots, is explicitly known \cite{DS}.
In terms of the Chebyshev polynomials of the first kind $T_{N}$,
defined by the identity $T_{N}(\cos\theta) = \cos N\theta$, we have
\be\label{solrankone} C(z) = 2 q^{1/2} T_{N}\Bigl(\frac{z}{2
q^{1/2N}}\Bigr)\, .\ee
Expanding in powers of $z$, we find an explicit parametrization of the
irreducible variety $\mathscr N_{|1)}$, generalizing \eqref{N3r1eq},
\be\label{sigmaform} \sigma_{2s+1} = 0\, ,\quad \sigma_{2s} =
(-1)^{s}\frac{N}{N-s}\binom{N-s}{s} q^{s/N}\, .\ee
Consistently with \eqref{vhatr}, $\mathscr N_{|1)}$ is an $N$-fold
cover of the $q$-plane. The roots of $C(z)$ can also be explicitly
computed,
\be\label{rootsnone} x_{i} = 2q^{1/2N}\cos\Bigl(\frac{\pi}{N}\bigl(
i-1/2\bigr)\Bigr)\, .\ee
In particular, one has $x_{i}=-x_{N+1-i}$ and, using
$\sigma_{2}=-Nq^{1/N}$ which follows from \eqref{sigmaform},
$x_{i}^{2} = -\frac{4}{N}\sigma_{2}\cos^{2}[\frac{\pi}{N}(i-1/2)]$. In
the case of even $N$, this implies that $C$ factorizes over $\mathbb
C(\mathscr N_{|1)})$ as
\be\label{Cfactrankone1} C(z) = \prod_{i=1}^{N/2}\Bigl( z^{2} +
\frac{4}{N}\sigma_{2}\cos^{2}\bigl(\frac{\pi}{N}(i-1/2)\bigr)\Bigr)\, .
\ee
In the case of odd $N$ we have similarly
\be\label{Cfactrankone2} C(z) = z\prod_{i=1}^{(N-1)/2}\Bigl( z^{2} +
\frac{4}{N}\sigma_{2}\cos^{2}\bigl(\frac{\pi}{N}(i-1/2)\bigr)\Bigr)\, .
\ee
From \eqref{Cfactrankone1} and \eqref{Cfactrankone2}, we immediately
read off the Galois group,
\be\label{GalNone} G_{|1)} = \mathbb Z_{2}\, .\ee
It is generated by the permutation $\tau$ acting on the roots as
\be\label{taugen}\tau (x_{i}) = x_{N+1-i}\, .\ee

The above examples show clearly that there is no direct relation
between the pattern of symmetry breaking in the usual classical sense
and the pattern of symmetry breaking described by Galois theory. This
should not be too surprising: the usual notion makes sense only
classically whereas the Galois symmetry makes sense in the quantum
theory; to a given phase may be associated several classical unbroken
gauge groups (examples are provided below) whereas the Galois symmetry
is a phase invariant.

\subsection{The Galois groups for $\text{U}(4)$}
\label{53s}

We have to compute the Galois groups of
\be\label{C4} C(z) = z^{4}+\sigma_{2}z^{2}-\sigma_{3}z+\sigma_{4}\ee
in the various phases of the model. The phases of rank four, three and
one have been studied in \ref{52s} and thus we can focus on the phases
of rank two. There are two such phases \cite{ferirr} that we shall
denote by $|2,t)$ with the confinement index $t=1$ or $2$.

\subsubsection{The group $G_{|2,2)}$}

The phase $|2,2)$ has confinement index $t=2$ with unbroken gauge
group $\text{U}(2)\times\text{U}(2)$. The irreducible variety
$\mathscr N_{|2,2)}$ is given by the condition that $C(z)+2q^{1/2}$
has two double roots. Taking into account the fact that the sum of all
the roots are zero, this implies that $C$ is of the form
\be\label{Cexpl20} C(z) = (z^{2}-a)^{2} - b\ee
which corresponds to the following parametrization of $\mathscr
N_{|2,2)}$,
\be\label{N422}\mathscr N_{|2,2)}:\ \sigma_{2} = -2 a\, ,\
\sigma_{3}=0\, ,\ \sigma_{4} = a^{2} -b\, ,\ q = b^{2}/4\, . \ee
One can invert these relations, $a=-\sigma_{2}/2$ and
$b=\sigma_{2}^{2}/4 - \sigma_{4}$, which shows that the variety
$\mathscr N_{|2,2)}$ is rational. Theorem \ref{monodGalois} then
implies that the Galois group $G_{|2,2)}$ coincides with the monodromy
group of the polynomial \eqref{Cexpl20}.

The discriminant of \eqref{Cexpl20} is given by
\be\label{discG22}\Delta = 256 b^{2} (a^{2}-b)\, .\ee
The monodromy group can be computed by considering the analytic
continuations of the roots along non-contractible loops on $\mathbb
C^{2}\setminus\Sigma$, where $\mathbb C^{2}=\{(a,b)\}$ and $\Sigma$ is
the zero locus of \eqref{discG22}. The roots of \eqref{Cexpl20} can be
easily written explicitly,
\be\label{rootsexpl} x_{1}=\sqrt{a+\sqrt{b}}\, ,\
x_{2}=\sqrt{a-\sqrt{b}}\, ,\ x_{3}=-\sqrt{a+\sqrt{b}}\, ,\
x_{4} = - \sqrt{a-\sqrt{b}}\, .\ee
The monodromy around $b=0$ yields the product of tranpositions
$(12)(34)$, the monodromy around $a=\sqrt{b}$ yields the transposition
$(24)$ and the monodromy arond $a=-\sqrt{b}$ yields the transposition
$(13)$. These permutations generate the symmetry group $\text{D}_{4}$
of a square whose vertices are labeled by the roots $x_{1},
x_{2},x_{3},x_{4}$ clockwise. The group $\text{D}_{4}$ is of order
eight and is called the dihedral group,
\be\label{group20} G_{|2,2)} = \text{D}_{4}\, .\ee
Let us note that the monodromies that one needs to consider in this
case can all be found by performing analytic continuations in the
semi-classical regime, which corresponds to small $q$ or equivalently
small $b$. This possibility corresponds to the fact that all the vacua
in the phase $|2,2)$ have the same unbroken gauge group and can all be
smoothly connected at weak coupling.

Another way to derive \eqref{group20} is to note that the polynomial
\eqref{Cexpl20} is a generic even polynomial. For a particular
ordering of the roots, consistent with the labeling \eqref{rootsexpl},
we thus have the relations
\be\label{rootrelC4} x_{1}+x_{3} = 0\, ,\quad x_{2}+x_{4} = 0\, .\ee
The only constraint on the Galois group is that it must preserve these
relations. In particular, the cycle $(1234)$ and the product of
transpositions $(12)(34)$ are in the group and these permutations
generate $\text{D}_{4}$. The only subgroup of $\text{S}_{4}$
containing strictly $\text{D}_{4}$ is $\text{S}_{4}$ itself. However,
the Galois group cannot be $\text{S}_{4}$ since, for example, the
permutation $(12)$ does not preserve the relations \eqref{rootrelC4}.
We conclude.

\subsubsection{The group $G_{|2,1)}$}

The phase $|2,1)$ contains vacua with either
$\text{U}(2)\times\text{U}(2)$ or $\text{U}(1)\times\text{U}(3)$
unbroken gauge groups \cite{csw1,fer2,ferirr}. The associated variety
$\mathscr N_{|2,1)}$, which is given by the vanishing of the
discriminants of $C+2q^{1/2}$ and $C-2q^{1/2}$, turns out to be
rational. As in the previous subsection, this will drastically
simplifies the calculation of the Galois group.

\begin{lem}\label{lemma22} The variety $\mathscr N_{|2,1)}$ is
rational, the birational mapping to $\mathbb C^{2}$ being given by
\begin{align}\label{ppp01} \sigma_{2} &= 2 (b-a^{2})\\\label{ppp02}
\sigma_{3} &= 4 ab\\\label{ppp03} \sigma_{4} &= \frac{1}{2}\bigl(
a^{4} - 6 a^{2}b + 2 b^{2}\bigr)\\\label{ppp04} q &=
\frac{1}{16}a^{2}\bigl( a^{2}- 4 b\bigr)^{3}\, .
\end{align}
\end{lem}

To prove this result, we write
\begin{align}\label{Cdec1} C(z) -
2q^{1/2}&=\bigl(z+(z_{1}+z_{2})/2\bigr)^{2}\bigl(z-z_{1}\bigr)\bigl(
z-z_{2}\bigr)\, ,\\
\label{Cdec2} C(z) -
2q^{1/2}&=\bigl(z+(z_{3}+z_{4})/2\bigr)^{2}\bigl(z-z_{3}\bigr)\bigl(
z-z_{4}\bigr)\, .
\end{align}
This yields
\begin{align}\label{aa1}\sigma_{2} &=
\frac{3}{4}\bigl(z_{1}^{2}+z_{2}^{2}\bigr)-\frac{1}{2}z_{1}z_{2}=
\frac{3}{4}\bigl(z_{3}^{2}+z_{4}^{2}\bigr)-\frac{1}{2}z_{3}z_{4}\, ,\\
\label{aa2}
\sigma_{3} &= \frac{1}{4}(z_{1}+z_{2})(z_{1}-z_{2})^{2} = 
\frac{1}{4}(z_{3}+z_{4})(z_{3}-z_{4})^{2}\, ,\\ \label{aa3}
\sigma_{4} &= \frac{1}{4}z_{1}z_{2}(z_{1}+z_{2})^{2} + 2q^{1/2} = 
\frac{1}{4}z_{3}z_{4}(z_{3}+z_{4})^{2} - 2q^{1/2}\, .
\end{align}
The idea is then to eliminate the variables $z_{1}-z_{2}$ and
$z_{3}-z_{4}$ from the above equations, keeping only $z_{1}+z_{2}$ and
$z_{3}+z_{4}$. After a straightforward but slightly tedious calculation
one obtains
\be\label{hhh} (z_{1}-z_{2})^{2} =
2(z_{3}+z_{4})(z_{1}+z_{2}+z_{3}+z_{4})\, ,\ 
(z_{3}-z_{4})^{2} =
2(z_{1}+z_{2})(z_{1}+z_{2}+z_{3}+z_{4})\, .\ee
Plugging this result into \eqref{aa1}, \eqref{aa2} and \eqref{aa3} and
defining
\be\label{abdef} a=\frac{1}{2}\bigl(z_{1}+z_{2}+z_{3}+z_{4}\bigr)\, ,\
b=\frac{1}{4}\bigl(z_{1}+z_{2}\bigr)\bigl(z_{3}+z_{4}\bigr)\, ,\ee
we get the parametrization given in the Lemma \ref{lemma22}. To prove
that the mapping is birational, we also have to express $a$ and $b$ in
terms of the $\sigma_{i}$s and $q$. Again this is a bit tedious but
straightforward to do and we find
\begin{align}\label{inv1} a &=\frac{4}{\sigma_{3}}\,
\frac{\sigma_{2}^5-12 \sigma_{4}
\sigma_{2}^3+6 \sigma_{3}^2 \sigma_{2}^2+32 \sigma_{4}^2 \sigma_{2}-18
\sigma_{3}^2 \sigma_{4}}{8 \sigma_{2}^3+27
\sigma_{3}^2}\, ,\\
\label{inv2} b &= \frac{1}{2}\sigma_{2} + a^{2}\, .
\end{align}

Combining Lemma\ \ref{lemma22} with Theorem \ref{monodGalois} in
Section \ref{monoSec}, we deduce that the Galois group $G_{|2,1)}$ in
the phase $|2,1)$ can be derived by computing the monodromy group of
\be\label{Cab} C(z)=z^{4} + 2\bigl(b-a^{2}\bigr)z^{2} - 4 a b z
+\frac{1}{2} \bigl(a^{4} - 6 a^{2}b + 2 b^{2}\bigr)\, .\ee
The discriminant of this polynomial is given by
\be\label{discr} \Delta = 32 a^{2}\bigl(a^{2} -
4b\bigr)^{3}\bigl(a^{4} + 10 a^{2} b -2 b^{2}\bigr)\, .\ee
If $\Sigma$ is the zero locus of $\Delta$, we have to consider
analytic continuations of the roots of $C$ along non-contractible
loops on $\mathbb C^{2}\setminus\Sigma$. We can simplify the analysis 
by noting that \eqref{Cab} is homogeneous, with $z$, $a$
and $b$ of degree one, one and two respectively. Since $b=0$ is not in
$\Sigma$, it is then convenient to rescale the variables in such a way
that $b=1$. We thus focus on
\be\label{Cb} C(z)=z^{4} + 2\bigl(1-a^{2}\bigr)z^{2} - 4 a z
+\frac{1}{2} \bigl(a^{4} - 6 a^{2} + 2\bigr)\, ,\ee
with 
\be\label{discrb} \Delta = 32 a^{2}\bigl(a^{2} -
4\bigr)^{3}\bigl(a^{4} + 10 a^{2}  -2 \bigr)\, .\ee

The zero locus $\Sigma$ of $\Delta$ corresponds to the points
$a=a_{i}$ and $a=-a_{i}$ with
\be\label{Sigmazero} a_{1}= 0\, ,\ a_{2}=2\, ,\ a_{3} =
\sqrt{-5+3\sqrt{3}}\, ,\ a_{4}=i\sqrt{5+3\sqrt{3}}\, .\ee
Let us pick a base point $a=a_{*}$ on the $a$-plane deprived of
$\Sigma$. For example, we can choose $a_{*}=i$. The roots of $C$ at
the base point are then given by
\be\label{rC} x_{1}\simeq -0.9 + 1.5 i\, ,\ x_{2} \simeq 0.9 + 1.5 i\,
,\ x_{3}\simeq -0.7 i\, ,\ x_{4}\simeq -2.2 i\, .\ee
A set of generators of the monodromy group is obtained by studying the
analytic continuations of the roots \eqref{rC} along contours
$\mathcal C_{i}$ and $\tilde{\mathcal C_{i}}$ that start from $a_{*}$
and circle counterclockwise around the points $a_{i}$ and $-a_{i}$
respectively (and not around any other point in $\Sigma$). Since $C$
is of degree four, this can be done from the exact formulas for the
roots. A more convenient method, that can be generalized to cases with
$N>4$, is to compute numerically the roots along the various contours
and read off the corresponding permutations on the data. This can be
easily implemented in Mathematica and we have used this method in the
following.

Taking into account \eqref{ppp04}, the monodromies around $a=0$ and
$a=\pm 2$ are at weak coupling. When $a=a_{1}=0$, $C(z) =
(1+z^{2})^{2}$ has two double roots. This case thus corresponds to the
$\text{U}(2)\times\text{U}(2)$ vacua. The monodromy around $\mathcal
C_{1}$ yields the permutation $(12)(34)$. This is reminiscent of
\eqref{taugen}, with a simultaneous exchange of the roots in the two
$\text{U}(2)$ factors. When $a= a_{2}= 2$, $C(z) = (z- 3)(z+ 1)^{3}$.
This corresponds to the weak coupling region with unbroken gauge group
$\text{U}(1)\times\text{U}(3)$. The monodromy around $\mathcal C_{2}$
must thus yield a transposition of the form \eqref{taugen}. Which
transposition it corresponds to precisely, taking into account that
the monodromy around $C_{1}$ is the permutation $(12)(34)$, involves
some strong coupling information. We find numerically the
transposition $(14)$. Similarly we have $\text{U}(1)\times\text{U}(3)$
vacua at $a=-a_{2}=-2$, since $C(z)=(z+3)(z-1)^{3}$. The associated
monodromy must be a transposition. Which one precisely involves again
some strong coupling effects and is found to be $(24)$. The
permutations $(12)(34)$, $(14)$ and $(24)$ generate the full symmetric
group $\text{S}_{4}$ and thus
\be\label{groupfinal} G_{|2,1)} = \text{S}_{4}\, ,\ee
without the need to consider additional monodromies.

Let us note that the same methods can be used to compute the Galois
groups in $\text{U}(N)$ theories for $N>4$. For example, in the
$\text{U}(5)$ theory, there are two phases $|1)$ and $|2)$ at rank
three and confinement index one. The associated varieties $\mathscr
N_{|1)}$ and $\mathscr N_{|2)}$ turn out to be rational. We have shown
that the Galois groups of both phases are $\text{S}_{5}$.
Unfortunately, in this case, the Galois symmetry does not distinguish
the phases.

\section{Conclusion}
\label{concS}

In this paper, we have explained how new types of symmetries are
hidden in the algebra of local observables in gauge theories. The
general idea is to reconstruct the gauge-variant data from the gauge
invariant observables by studying the gauge invariant set of equations
that the gauge-variant quantities must satisfy. The solution of this
problem is governed by a Galois symmetry group, which may be a strict
subgroup of the gauge group. Remarkably, this provides a perfectly
well-defined notion of gauge symmetry breaking at the quantum level.

In particular, the Galois symmetry is a phase invariant and can thus
be used to classify the phases, unlike the usual pattern of gauge
symmetry breaking which can be different in two vacua belonging to the
same phase. We have been able to compute explicitly a few Galois
groups using elementary methods. The classification obtained in terms
of the Galois groups studied in this paper does not provide a complete
classification of the phases, as explained at the end of the previous
Section. Our analysis was restricted to the $\text{S}_{N}$ subgroup of
the full gauge group and it is conceivable that a finer classification
could be achieved by using the full gauge symmetry along the lines
sketched in Section \ref{generaSec}.

It is natural to ask whether the Galois symmetries could play a r\^ole
in other related problems, like for example to better understand
't~Hooft's monopoles in the abelian projection or the Gribov
ambiguity.\footnote{I would like to thank M.~Dougals for pointing out
this possibility.} It is also interesting to note that in quantum
gravity, the lack of local observables is very similar to the lack of
local order parameters in gauge theory. This suggests that Galois
theory might be useful in this context as well.

\subsection*{Acknowledgements}

This work is supported in part by the belgian Fonds de la Recherche
Fondamentale Collective (grant 2.4655.07), the belgian Institut
Interuniversitaire des Sciences Nucl\'eaires (grant 4.4505.86) and the
Interuniversity Attraction Poles Programme (Belgian Science Policy).
The author is on leave of absence from Centre National de la Recherche
Scientifique, Laboratoire de Physique Th\'eorique de l'\'Ecole Normale
Sup\'erieure, Paris, France.

\renewcommand{\thesection}{\Alph{section}}
\renewcommand{\thesubsection}{\arabic{subsection}}
\renewcommand{\theequation}{A.\arabic{equation}}
\setcounter{section}{0}
\end{document}